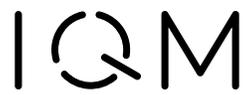

# Technology and Performance Benchmarks of IQM's 20-Qubit Quantum Computer

IQM Garnet

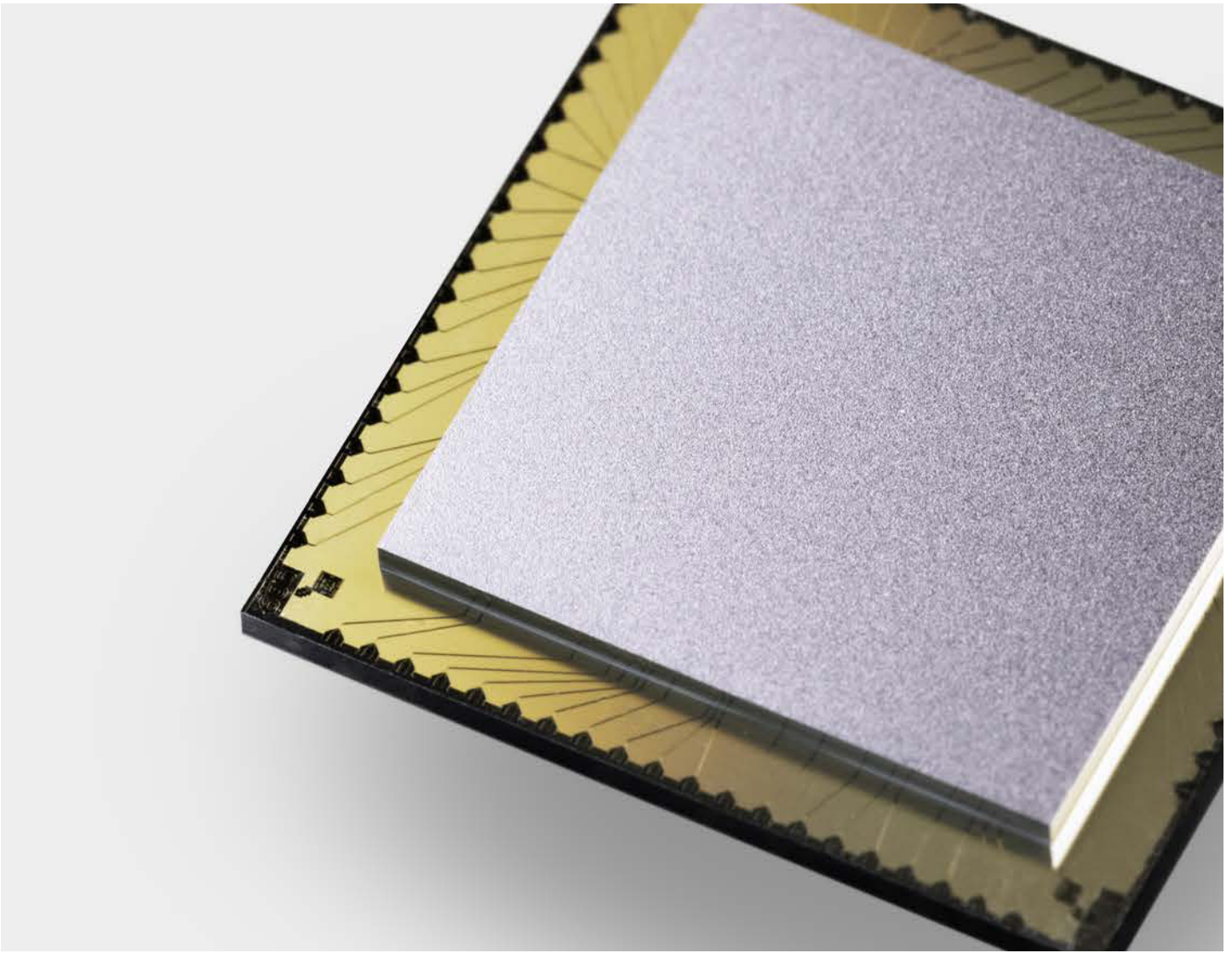

# Contributors

Leonid Abdurakhimov, Janos Adam, Hasnain Ahmad, Olli Ahonen, Manuel Algaba, Guillermo Alonso, Ville Bergholm, Rohit Beriwal, Matthias Beuerle, Clinton Bockstiegel, Alessio Calzona, Chun Fai Chan, Daniele Cucurachi, Saga Dahl, Rakhim Davletkaliyev, Olexiy Fedorets, Alejandro Gomez Frieiro, Zheming Gao, Johan Guldmyr, Andrew Guthrie, Juha Hassel, Hermanni Heimonen, Johannes Heinsoo, Tuukka Hiltunen, Keiran Holland, Juho Hotari, Hao Hsu, Antti Huhtala, Eric Hyyppä, Aleksi Hämäläinen, Joni Ikonen, Sinan Inel, David Janzso, Teemu Jaakkola, Mate Jenei, Shan Jolin, Kristinn Juliusson, Jaakko Jussila, Shabeeb Khalid, Seung-Goo Kim, Miikka Koistinen, Roope Kokkoniemi, Anton Komlev, Caspar Ockeloen-Korppi, Otto Koskinen, Janne Kotilahti, Toivo Kuisma, Vladimir Kukushkin, Kari Kumpulainen, Ilari Kuronen, Joonas Kylmälä, Niclas Lamponen, Julia Lamprich, Alessandro Landra, Martin Leib, Tianyi Li, Per Liebermann, Aleksi Lintunen, Wei Liu, Jürgen Luus, Fabian Marxer, Arianne Meijer-van de Griend, Kunal Mitra, Jalil Khatibi Moqadam, Jakub Mrożek, Henrikki Mäkynen, Janne Mäntylä, Tiina Naaranoja, Francesco Nappi, Janne Niemi, Lucas Ortega, Mario Palma, Miha Papič, Matti Partanen, Jari Penttilä, Alexander Plyushch, Wei Qiu, Aniket Rath, Kari Repo, Tomi Riipinen, Jussi Ritvas, Pedro Figueroa Romero, Jarkko Ruoho, Jukka Räbinä, Sampo Saarinen, Indrajeet Sagar, Hayk Sargsyan, Matthew Sarsby, Niko Savola, Mykhailo Savytskyi, Ville Selinmaa, Pavel Smirnov, Marco Marín Suárez, Linus Sundström, Sandra Słupińska, Eelis Takala, Ivan Takmakov, Brian Tarasinski, Manish Thapa, Jukka Tiainen, Francesca Tosto, Jani Tuorila, Carlos Valenzuela, David Vasey, Edwin Vehmaanperä, Antti Vepsäläinen, Aapo Vienamo, Panu Vesanen, Alpo Välimaa, Jaap Wesdorp, Nicola Wurz, Elisabeth Wybo, Lily Yang and Ali Yurtalan

Quantum computing has tremendous potential to overcome some of the fundamental limitations present in classical information processing. Yet, today's technological limitations in the quality and scaling prevent exploiting its full potential. Quantum computing based on superconducting quantum processing units (QPUs) is among the most promising approaches towards practical quantum advantage.

In this article the basic technological approach of IQM Quantum Computers is described covering both the QPU and the rest of the full-stack quantum computer. In particular, the focus is on a 20-qubit quantum computer featuring the IQM Garnet QPU and its architecture, which we will scale up to 150 qubits. We also present QPU and system-level benchmarks, including a median 2-qubit gate fidelity of 99.5% and genuinely entangling all 20 qubits in a Greenberger-Horne-Zeilinger (GHZ) state.

# 1  Introduction

The exponential nature of the quantum state, represented by the complex valued $2^N$-dimensional state vector of a quantum computer with $N$ qubits leads to the failure of any classical simulation to predict the operation of even moderate-sized systems. For the most powerful classical supercomputing clusters, the limit is today on the order of 50 qubits, above which the simulation becomes unfeasible in humanly accessible computing times [2, 50]. There are many known quantum algorithms today that can exploit the properties of quantum mechanics to overcome some fundamental limitations of classical information processing, enabling computations otherwise impossible or extremely hard [31]. Yet, the immature state of quantum hardware today prevents full exploitation of the potential.

Many variants of superconducting qubits and quantum processors have been introduced to date [21]. In this white paper we present the solutions developed by IQM. The qubits and coupling structures of the QPU are described along with other subsystems, including the cryogenic system with connectivity, room temperature control electronics, and software. These solutions are employed in IQM systems with up to 150 qubits. High-quality quantum computing systems of this scale have potential for early quantum utility, and they constitute a necessary milestone in the roadmap towards broader quantum advantage. In more detail, we describe the 20-qubit quantum computer representative of IQM core technology choices. To demonstrate the performance enabled by this technology, we present extensive benchmarking results from foundational level fidelity benchmarks to application benchmarks quantifying performance in selected classes of computational tasks.

# 2  QPU Architecture

IQM offers QPUs with different architectures. While resonator star architecture provides higher connectivity [20, 1], qubit crystal topology allows more parallelism, further discussed below. Today, IQM offers qubit crystals with computational qubit count from 5 to 150, see Fig. 1, with the focus of this white paper being on the 20-qubit QPU we refer to as IQM Garnet. In IQM qubit crystals, qubits are arranged on a square lattice where the lattice is ro-tated respect to crystal edge by 45 degrees. Qubit-qubit connectivity is mediated by tunable couplers. Both computational qubits and tunable couplers are based on flux-tunable transmon qubits. Qubit states are probed by frequency-multiplexed dispersive readout.

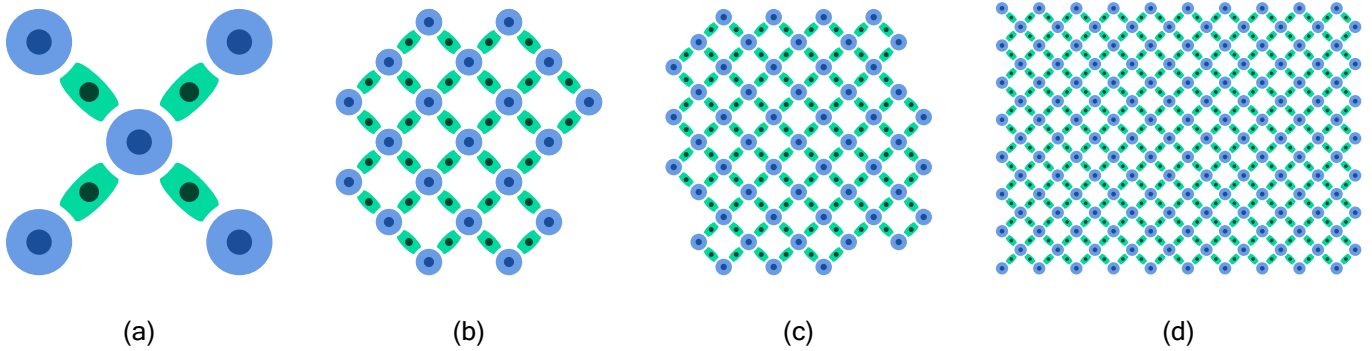

Figure 1: IQM qubit crystal QPU topologies (a) with 5, (b) with 20, (c) with 54, (d) and with 150 computational qubits (blue) interconnected with tunable couplers (green).

## 2.1 Connectivity

The main way in which gate errors influence the final error of the algorithm is via the algorithm's runtime; in particular, its dependence of the circuit's depth. Within planar geometries, the square lattice with nearest-neighbor connectivity is close to ideal as a general purpose platform providing a balance between parallelism and worst-case routing distance. Consider an $N$-qubit circuit with commuting 2-qubit gates between each and every qubit, for example the phase separator in a QAOA circuit for a Sherrington-Kirkpatrick spin-glass problem. On an all-to-all connected QPU with no parallelization capabilities the runtime would be proportional to $N^2$, while for the fully parallelized one-dimensional QPU the runtime would be proportional to $N$ [49]. The square grid QPU has the same scaling as the one-dimensional QPU with a better prefactor. In contrast, for a sparse circuit, where only few gates can run in parallel due to the nature of the algorithm, like QAOA for MaxCUT on regular graphs, the circuit runtime scales as $\sqrt{N}$ [42, 24] outperforming maximally parallel qubit chain.

The mentioned strengths of square lattice also reduce the gate performance requirement for quantum supremacy demonstrations using random circuit sampling experiments [2] because of the increased tensor-network simulation complexity of this topology compared to, for example a heavy-hex topology [8, 44].

Furthermore, the square array topology is compliant with standard surface code based error correction schemes. For surface code experiments, we have further increased the utility by the choice of angle in between lattice and crystal edge which increases the number of possible weight-4 parity checks for the given qubit number.

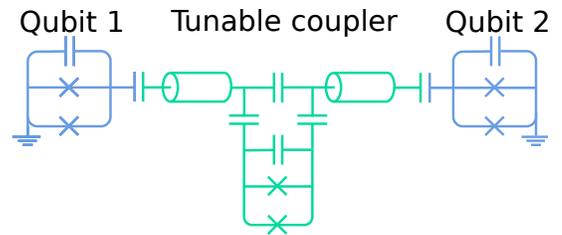

Figure 2: Circuit representation of the qubit-coupler-qubit unit

## 2.2 Quantum gates

The QPU and control system architecture support a universal gate-set based on single-qubit rotation gates around X and Y axes, arbitrary-angle virtual-Z gate [30], and the conditional phase gate (CZ). The single-qubit rotation gates are implemented as microwave pulses applied through control lines that are capacitively coupled to the qubit island. The pulse shapes are optimized to minimize the leakage to any of the adjacent qubit states or to the higher excited states of the driven qubit [16].

The qubits are coupled to each other through transmon-based couplers that enable strong tunable $ZZ$-type interaction up to $50$ MHz which can be fully turned off during the idling operation [29]. In contrast, static capacitive coupling would limit the effective qubit-qubit interaction on-off ratio to about 50, resulting in either slow gates or unacceptable amount of $ZZ$-type errors. The CZ-gate is implemented by applying a magnetic flux pulse on the coupler, resulting in a fast gate with a typical duration of $20$ ns - $40$ ns.

In addition to the CZ gate, the architecture supports multiple other gate types, such as the cross-resonance gate [40], the parametric resonance gate [39], or the iSWAP gate [43], which enables choosing the optimal gate-set to fit the particular purpose. In addition, the qubit frequencies can be tuned by current-biasing magnetic flux lines of individual qubits, enabling efficient execution of analog quantum simulations on the QPU [19].

## 2.3 Qubit readout

For qubit readout, each qubit has its own readout structure composed of a narrow band readout resonator and a wide band Purcell filter [14], with the former being capacitively coupled to the qubit island. A subset of readout structures is then coupled to a wider passband-like filter embedded in the probe line. The standing wave in the readout probe line enhances the coupling with readout structures and enables faster qubit readout. The variant of IQM Garnet presented in this whitepaper features 3 probe lines with 7, 7 and 6 readout structures coupled to each, see the example spectrum in Fig. 3. Although the transmission spectrum of readout structures is non-trivial, standard qubit state discrimination methods can be used and this architecture has been shown to have low readout crosstalk [14].

## 2.4 Crosstalk

A key aspect of any scalable architecture is the suppression of crosstalk to enable parallel operation. Above, we already mentioned reduction of readout crosstalk using individual Purcell filters and cancellation of $ZZ$-interactions between idled neighboring qubits using the tunable couplers. In addition, our tunable coupler design, see Fig. 2, features coupling extenders which feature compact field distribution to reduce qubit drive crosstalk and provide space in the qubit lattice for aforementioned readout structures [29]. Also, the specific topology of the tunable coupler allows them to be operated at a large detuning from the readout structures reducing readout crosstalk further and improving gate fidelity.

## 2.5 Control routing and fabrication

The drawback of the nearest-neighbor connectivity implemented with tunable couplers is 2.5x increase in the number of tunable transmons from $20$ computational qubits to $50$ tunable transmons in total. Moreover, each tunable coupler requires a control line for the fast magnetic flux pulses. Adding to the coupler control lines the aforementioned qubit drive, flux and readout lines, IQM Garnet is controlled by 76 control lines, or 3.8 lines per qubit. The requirements for the fabrication technology and system-level complexity is essentially set by the number of physical qubits. The fabrication complexity is addressed by 3D integrated stack featuring routing and qubit chips connected through a superconducting flip-chip technology. IQM QPUs are fabricated in IQM's dedicated fabrication facility in Espoo, Finland capable of high-yield production of superconducting quantum devices [25, 26].

# 3 System

## 3.1 General hardware description

The quantum computer is designed with practical considerations, prioritizing features such as noise reduction, minimal vibrations, ease of installation, and efficient deliverability. Internal components like water distribution, power supply, networking, and compressed air facilities are enclosed within the system. The system features a Bluefors XLD dilution refrigerator as the cryogenic host, supporting the QPU with a standard cascade of attenuation and filtering components.

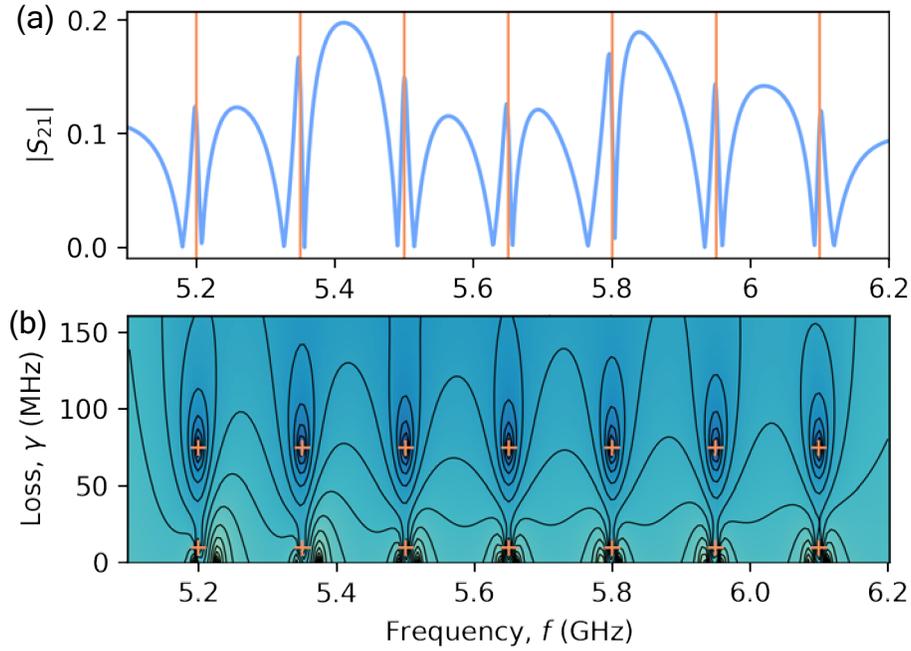

Figure 3: Designed readout circuit (a) transmission magnitude as a function of frequency $S_{21}(f)$ and (b) impedance magnitude $\log_{10}(|Z_{11}|)$ as a function of complex frequency $s$, where red crosses indicate target mode locations in frequency and linewidth.

Magnetic shields are in place to protect critical components including the QPU and Traveling Wave Parametric Amplifier (TWPA) [35]. Microwave isolators are further shielded and strategically located above the mixing chamber flange to minimize interference, Fig. 4 shows the cryogenic components. Other aspects of the cryogenic design implement the known best standards to use high purity copper, non-magnetic gold plating, intermediate thermalization in the filter stack, and tidy cable clamping.

The entire system is arranged to fit with a standard row of equipment racking used typically by High Performance Computing (HPC) centers, seen in Fig. 5, and includes itself a standard 19" electronics rack. The measurement rack (Fig. 5a) hosts the IQM Quantum Control System and includes a double conversion Uninterruptible Power Supply (UPS) for a reliable and electrically cleaner power supply. An additional auxiliary electronics rack handles system monitoring, fridge control, and provides a local access terminal for manual control.

To minimize physical presence requirement, the system incorporates remote power management, monitoring, and control of system fans. Secondary access administration channels are integrated for remote management of the servers without physical intervention. Network security is ensured through fully managed network switches and a firewall.

## 3.2 QPU control electronics

The microwave pulses for single-qubit rotations, the base-band flux pulses for two-qubit gates, and the microwave probe pulses for readout of the qubits are generated by the in-house designed IQM Quantum Control System (QCS). The system is built around PXIe infrastructure which provides modularity, scalability, and high-speed connections between the modules, see Fig. 6. The microwave pulses for single-qubit rotations are generated by direct digital synthesis, allowing a wide bandwidth and avoiding the complexities related to mixer-based solutions. The DC-coupled modules for generating the flux signals include two separate digital-to-analog converters specifically chosen to provide both stable DC bias signals and pulses with a sufficient bandwidth for a rapid and well-controlled ramp-up and ramp-down. The readout module is designed to generate frequency multiplexed probe pulses and process the response for simultaneous readout of up to ten qubits. The real-time control of the executed pulse sequence is implemented with a field-programmable gate array (FPGA) associated with each module of the system. Additionally, IQM QCS implements an advanced clock distribution and timing solution, enabling accurate phase and event synchronization across all the channels in the system. Overall, the IQM QCS has been designed to provide high-quality control signals specific to the chosen QPU architecture in a cost-effective manner.

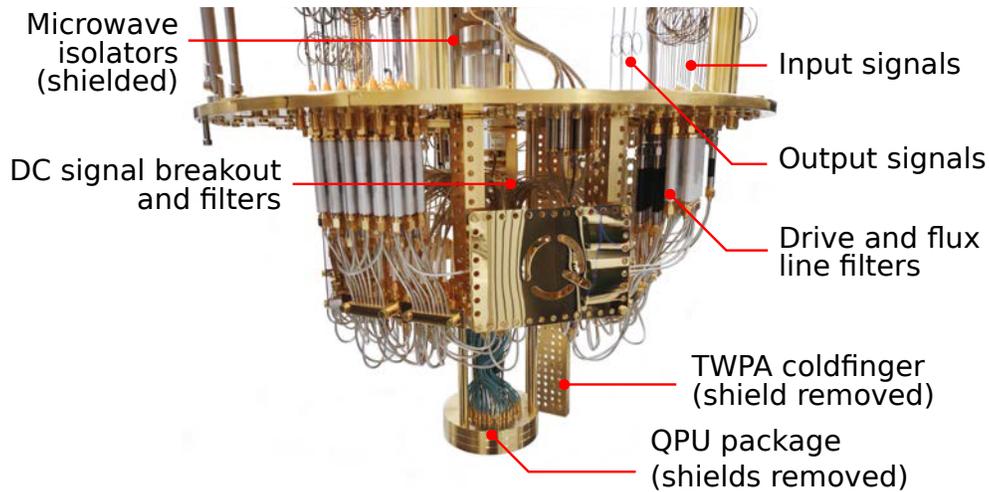

Figure 4: Cryogenic components inside the host dilution refrigerator

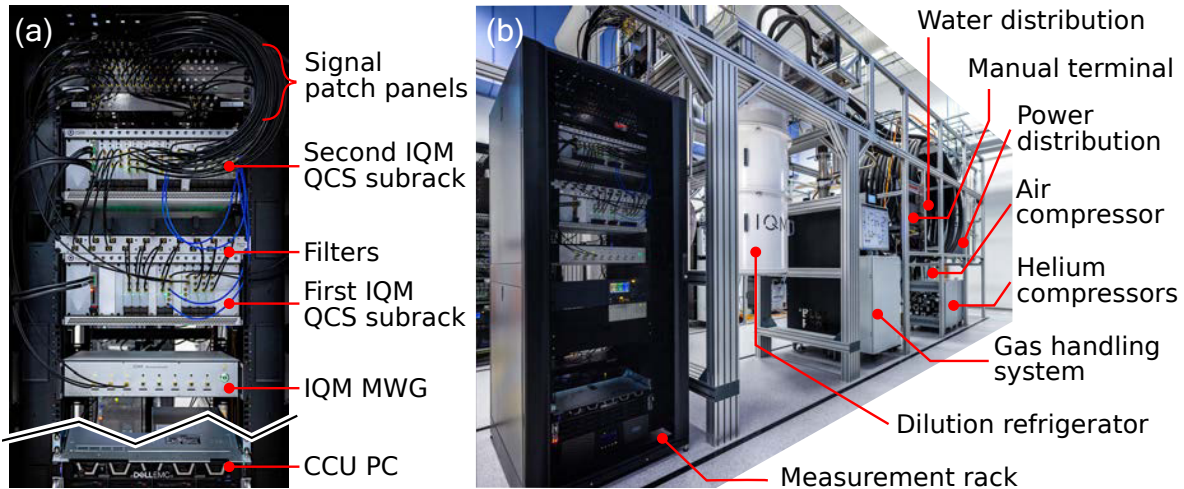

Figure 5: Photograph of the system without covers

In addition to the IQM QCS, the control electronics setup includes an in-house designed IQM Microwave Generator (MWG) for the continuous wave pump signals for the TWPA, and a Qblox D5a module for DC flux currents for qubits which are operated at fixed frequency throughout computation.

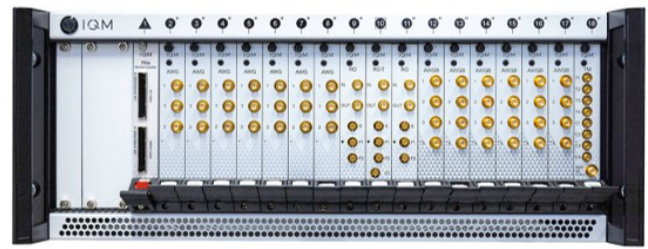

Figure 6: Photograph of a fully populated subrack of IQM QCS electronics.

### 3.3 QC control software

The control software stack is divided into multiple functional layers presented in Fig. 7. It can be configured based on the required level of access. The software modules and their roles are described in more detail in the following subsections.

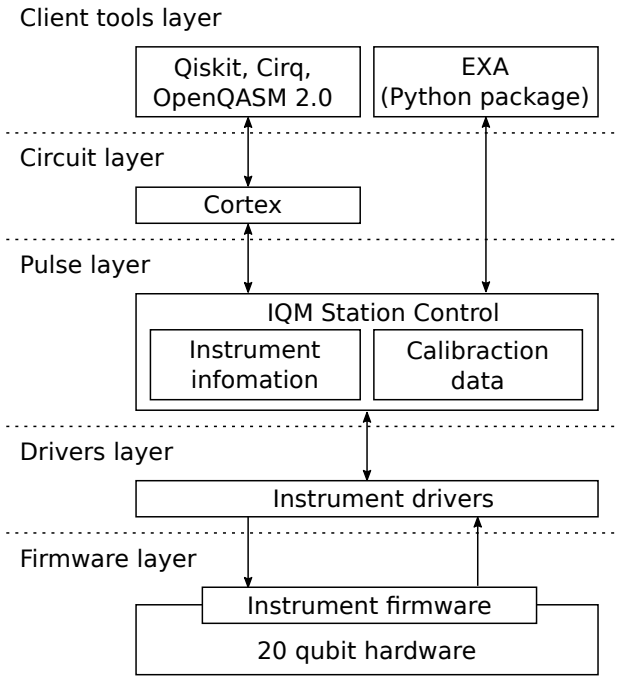

Figure 7: The software layers and modules of our quantum computer control software stack.

IQM QCCSW is delivered as a Debian package, enveloping application images, helper scripts and configuration files, and a `systemd` service. By packaging the images and relevant files with Debian package manager, installation and up-keeping become easy, offering easy upgrades and configuration management. Containerisation brings in capabilities like resource allocation and consolidating dependency. Coupled with robust service management capabilities like boot process optimisation, service isolation, and automatic restarts, the software stack ensures reliability, security and resilience.

### 3.3.1 Cortex

Cortex is the highest level of abstraction in the QCCSW stack, that allows running quantum algorithms on IQM quantum computer. It is built to enable quantum computation for the end user, rather than experimenting with the behaviour of underlying elements of the quantum computer itself. Cortex users can define and execute the quantum algorithms, expressed as quantum circuits using application-level frameworks like Qiskit, Cirq and OpenQASM 2.0. The input to the IQM quantum computer is a computational job with one or more quantum circuits, the shot count, and any possible parameters defined by the client tool. The job is then queued for execution and the results of the circuit measurements can be retrieved when the job is completed.

### 3.3.2 EXA

EXA is a comprehensive, customisable Python library for characterisation, calibration and control of IQM quantum computers. The central feature of EXA is the Experiment class, which gives the user pulse-level access to the quantum computer and combines functionalities such as waveform control, execution flow, data manipulation, analysis and presentation. It allows the user to change any parameter of the system and defines sensible defaults which can be easily expanded or overridden. The EXA Experiment Library enables calibration of quantum computers, greatly simplifying measurement and control processes. It helps automate standard procedures, reduce repetitive tasks, and develop entirely new experiments with only minimal amounts of new code.

### 3.3.3 IQM Station Control

IQM Station Control is responsible for low-level functionalities like instrument parameters, hardware drivers and persistence of raw data and metadata. It abstracts out the hardware details from higher-level components such as EXA and Cortex.

Station Control service's non-RESTful JSON (JavaScript Object Notation) HTTP (HyperText Transfer Protocol) interface handles communication with Cortex and EXA. Parameter sweeps and pulse schedule executions can be carried out using the endpoints provided by this interface. The user does not need to directly interact with the service normally. Device-specific drivers help Station Control to encapsulate the details of each instrument, including its low-level communication protocol.

# 4 Performance benchmarking

As the name suggests, quantum computers are built for practical computing purposes. As practical quantum advantage is yet to be achieved, and it is unclear where it will first be discovered, the field measures quantum computer performance through a variety of benchmarking metrics [22, 9]. At IQM we have taken a four-fold approach to benchmarking:

- Foundational operation level
- System total operation level
- Fundamental physics based
- Application benchmarks

With these benchmarks, we can understand the operation of the device on the component level (foundational operation level) while making sure we can track the performance of the device as a whole (system total operation level). The fundamental physics and application benchmarks are needed to ensure the metrics are relevant to end applications allowing for improving our quantum computers over time. We provide benchmarking results for IQM Garnet on all four levels below.

## 4.1 Foundational operation level benchmarks

High-performance quantum computing builds on low error rates of the gate-level operations. As discussed in Section 2.4, a key enabler for low QPU-level error rates in parallel operation is low crosstalk. We define flux and drive crosstalk by

$$C_{\text{Flux}} = 20 \log_{10} \frac{\partial \Phi_i / \partial I_j}{\partial \Phi_j / \partial I_j}, \tag{1}$$

$$C_{\text{Drive}} = 20 \log_{10} \frac{\partial \Omega_i / \partial V_j}{\partial \Omega_j / \partial V_j} \tag{2}$$

correspondingly where $\Phi_i$ is the flux bias, $I_i$ is the flux line current, $\Omega_i$ is the Rabi rate for resonant drive, and $V_i$ drive pulse amplitude applied on the drive line of qubit $i$. On IQM Garnet devices, we have measured median crosstalk level of $C_{\text{Flux}} = -70$ dB and $C_{\text{Drive}} = -48$ dB, see Fig. 8a. The observed median values for flux crosstalk are more than 3 dB, and for drive crosstalk are more than 6 dB lower compared to other recently published QPU with a similar architecture [23] and more favorably compared to most older designs and to other quantum computing platforms [23]. In a smaller demonstrator system with wirebond-free package 9 dB lower drive crosstalk is demonstrated, indicating possible future gain by upgrading to the next generation packaging solution [41].

We obtain average single qubit gate fidelity by averaging over Clifford group in a randomized benchmarking [10] experiment and normalize the error per native gate. For characterizing CZ gate error we use interleaved randomized benchmarking [27]. We characterize gates in parallel in distance-two groups, where qubits are separated by two idling couplers and one idling qubit and obtain median single qubit gate error of $9 \times 10^{-4}$, CZ error of $5 \times 10^{-3}$ and two qubit Clifford error of $2 \times 10^{-2}$, see Fig. 8b. Benchmarking the two qubit gates in distance-one groups results in a reduction of mean fidelity by $0.6$ for CZ gates and $3$ percentage points for two-qubit Clifford gates.

We define readout error as the probability of not discriminating the qubit to be in the prepared state without correcting for the state preparation errors [45]. We observe median readout error for simultaneous QPU readout to be $3 \times 10^{-2}$.

## 4.2 System level benchmarks

The system level benchmarks measure the join operation of several qubits or the whole processor in one go. They measure how the entire device operates, and provide information on the strengths and weaknesses of the specific calibration and parts of the quantum computer. Typically they do not however directly predict performance on practical tasks quantum computers will be used for. All the methods presented in this section are based on random quantum circuits.

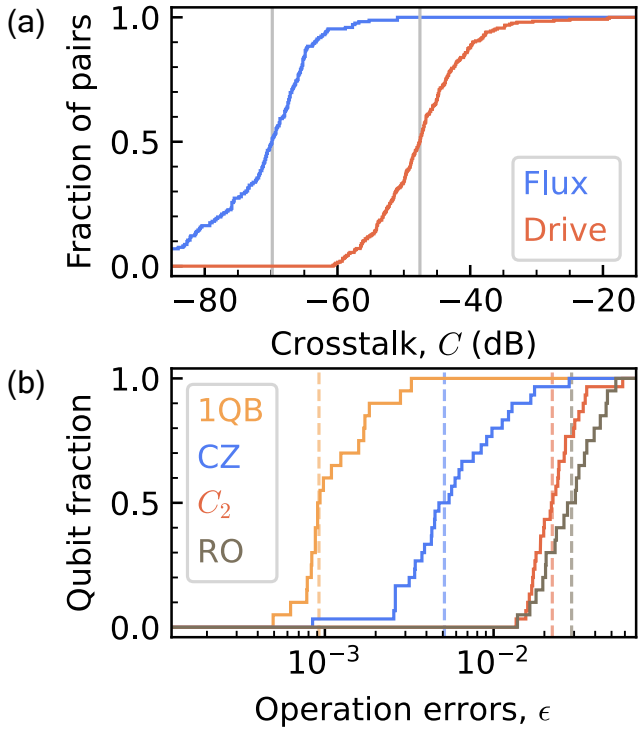

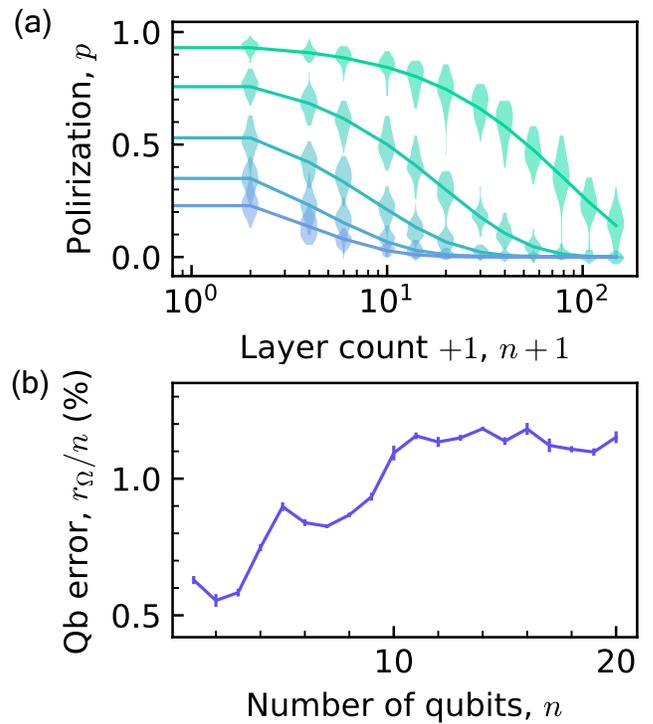

Figure 8: (a) Cumulative distribution of drive and flux control crosstalk between pairs of operation target and spectating qubits. (b) Cumulative distribution of average single-qubit gate (1QB), native two qubit gate (CZ), composite two-qubit Clifford operation ($C_2$), and readout (RO) operation errors for IQM Garnet QPU components, with vertical lines indicating median values.

Figure 9: (a) State polarization distributions (violins) after $d$ layers of random gates in MRB experiments together with exponential fits for $n = 2, 6, 10, 16, 20$ qubits (blue to green). (b) Error per gate layer per qubit extracted from the fitted decay constants like in the first subfigure.

### 4.2.1 Mirror randomized benchmarking

Randomized mirror circuits combine a mirrored structure with randomized compiling to enable scalable and robust randomized benchmarking (RB) of both Clifford and universal multi-qubit gate sets without the classical computation overhead [37, 15]. We carried out MRB protocol by interleaving layers with gatesets $\mathbb{G}_1 = \mathbb{C}_1$, where $\mathbb{C}_1$ is the one-qubit Clifford group, and $\mathbb{G}_2 = \{CZ\}$ with a uniform CZ-gate density $\xi = 1/2$ and a probability distribution $\Omega$ over an $n$-qubit layer set. We observe expected decay of state polarization with increasing gate layer count, see Fig. 9a. By fitting exponential decay models and comparing decay constants normalized by qubit count we see evidence of minor increase of error per layer per qubit $r_{\Omega, perQ} = 1 - (1 - r_\Omega)^{1/n} \approx r_\Omega/n$ with increased number of qubits, see Fig. 9b. This degradation is like due to the fact that larger benchmark has to in- clude qubit pairs with sub-median performance as well as some crosstalk effects.

### 4.2.2 Quantum Volume and volumetric benchmarking

To get a broader overview of the ability of the device to run circuits of various types and sizes, and to measure the joint scaling of fidelity and qubit count, we have performed volumetric benchmarking [5, 36] and measured the Quantum Volume of IQM Garnet [7]. Quantum volume (QV) is measured by performing "square shaped" circuits where the depth of the circuit equals the width (number of qubits), such that each layer contains a maximal number of random $SU(4)$ two-qubit unitaries between random pairs of qubits. Volumetric benchmarking framework was used here to measure the output fidelities of random circuits with a varying number of qubits and depths.

The result of a successful $QV = 2^5$ benchmark experiment is shown in Fig. 10a, with average heavy-output probabilities converging to a value greater than $2/3$ within two standard deviations. In Fig. 10b, we show the volumetric benchmark background of average fidelities of sets of 40 random circuits with layers of gates consisting of uniformly sampled single-qubit Clifford gates and CZ gates, with a CZ gate density of $\xi = 1/4$. The data surrounded by black borders are measured data, while the other boxes are extrapolated fidelities based on the measured data.

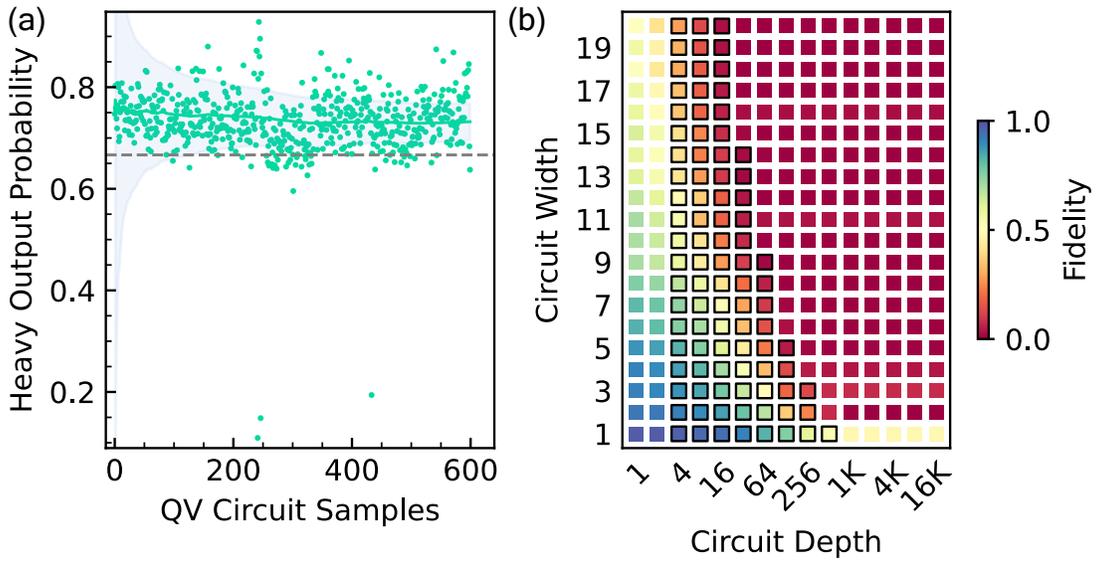

Figure 10: (a) Observed heavy-output probability for circuit samples for test size $QV = 2^5 = 32$. (b) Volumetric benchmarking fidelity using random circuits. The data surrounded by black borders are measured data, while the other boxes are extrapolated fidelities based on the measured data.

### 4.2.3 Circuit Layer Operations Per Second (CLOPS)

Circuit Layer Operations Per Second (CLOPS) measure the execution speed of the random circuits used to measure quantum volume of the device [46]. CLOPS also include a measure of the feedback time between the control computer and the quantum computer that is typical for variational algorithms by including parameter updates to the random circuits. The fast execution of circuits that is measured by CLOPS is crucially important for the usage of quantum computers, as a faster through-put rate enables applying more and heavier error-mitigation techniques. These are essential for high-quality computing and research results using current and near-term quantum computers. Recently there has been a new hardware-aware definition of CLOPS that only applies gates between qubits that are connected on the hardware [17], and in this context the quantity reported here is called the virtual CLOPS.

CLOPS is computed as

$$\text{CLOPS} = \frac{M \times K \times S \times D}{T} \quad (3)$$

where:

- $M$ = number of random circuit templates = 100
- $K$ = number of variational parameter updates = 10
- $S$ = number of shots = 100
- $D$ = number of QV layers = $\log_2 QV$
- $T$ = time taken

We have measured a CLOPS of $2600$ on IQM Garnet. The result reflects partially the quantum volume, as indicated by $D$ in the formula, but mainly the speed of the execution. Factors that determine it are control electronics, software and signal latency communicating between the quantum and classical computers. Section 3 describes the control electronics and software that enable the fast execution of quantum circuits on IQM Garnet.

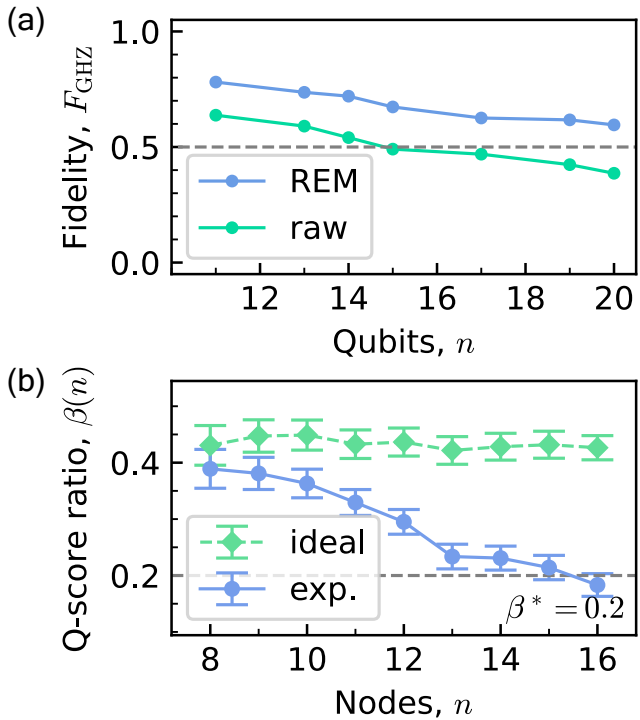

Figure 11: (a) GHZ-state fidelity $F_{\mathrm{GHZ}}$ across the number of entangled qubits with readout read-out-error mitigation (REM, blue), without REM (green) and fidelity threshold $F_{\mathrm{GHZ}} > 0.5$ for genuine multi-qubit entanglement (gray dashed). (b) Measured Q-score ratio $\beta(n)$ (blue) with one virtual node as a function of node count $n$ on $n-1$ qubits compared to threshold of $0.2$ (gray dashed), indicating Q-score of $15$, and expected outcome for ideal quantum computer (green).

## 4.3 Fundamental physics benchmarks

### 4.3.1 GHZ state creation

With the system-level benchmarks performed, it is time to start looking at the quantumness properties of IQM Garnet. A foundational requirement for useful quantum computing that has potential to go beyond classical capabilities is to entangle all qubits of the processor into a genuinely multi-qubit entangled (GME) state. The typical way to demonstrate GME is to prepare a GHZ state [12]. A GHZ state fidelity $F_{\mathrm{GHZ}} > 0.5$ is a witness for GME [48] and can be measured using the method of multiple quantum coherences [47, 3]. Fig. 11a reports GHZ state fidelities measured on IQM Garnet with and without readout-error mitigation (REM). The results without REM show that GME can be certified up to 14 qubits according to the strictest standards of fundamental physics. A quantum computer however typically does not need to prepare entangled states for the sake of it, but to use them as a resource in computation. In a computational scenario measurement is deferred to the end of the algorithm, so we are interested in the GHZ state fidelity pre-measurement. This is what we access by applying readout-error mitigation using the mthree package [33]. The readout-error mitigated results show that we have prepared a 20-qubit genuinely multi-qubit entangled GHZ state with fidelity $F_{\mathrm{GHZ20}} = 0.62$, verifying the non-classical nature of the entire IQM Garnet system.

## 4.4 Application benchmarks

### 4.4.1 Q-Score: solving Max-Cut with QAOA

Finally, we want to see how IQM Garnet performs in providing solutions to computational problems. To showcase performance in this arena we choose to show results from computing the solutions to the maximum-cut combinatorial optimisation problem using the Q-score benchmark [28]. IQM Garnet has been shown to pass the Q-score benchmark up to problem-size 15, as shown in 11b. Q-score is defined on random Erdős-Rényi graphs with edge probability $p = 0.5$ for each connection. The Q-score test is passed if $\beta(n) \geq 0.2$, where $\beta(n)$ measures the fraction of the optimality gap between a random guess and the optimal solution to the problem graph on $n$ qubits the quantum computers has captured. Due to the accumulation of noise, the Q-score test becomes increasingly hard with growing problem size.

We measure our Q-score performance with a depth $p = 1$ QAOA circuits [11, 4] execution, where we find the optimal circuit parameters by evaluating analytical formulas [34]. This method is enabled by the fact that local expectation values evaluated on the $p = 1$ QAOA state only depend on a reduced number of qubits. Such angle optimization procedure is classically efficient, and ensures the optimal use of quantum resources. We also increase the score by $1$ through use of the virtual node technique [6, 38] that introduces an overhead of only a few single-qubit $Z$-gates.

To further improve performance, we execute the quantum circuits on an optimal layout using the method introduced in [32]. Here, in a pre-processing step, we optimise a noise score for each of the layouts based on the gates count of the circuit and the single- and two-qubit-gate error rates of the hardware. We use readout-error mitigation implemented in the mthree package [33], which allows us to leave out readout errors from the noise score.

## 5 Summary and outlook

We presented a versatile set of benchmarking data for a quantum computing system built on IQM Garnet, a 20-qubit quantum processor with square lattice topology and a dedicated tunable coupler solution for high-quality two-qubit gate operations. The basic technical solutions are representative of IQM's qubit crystal QPU family. The five-qubit QPU is the core of the IQM Spark system forming a low-cost on-premises product for research and education use [38]. IQM qubit crystals, see Sec. 2, are the QPUs of the IQM Radiance integrated quantum computing systems, representing the current state of art in commercially available on-premises superconducting quantum computers. These systems are useful for exploring the limits of quantum computation, including use cases approaching quantum utility and advantage, and application in the context of quantum acceleration for classical high-performance computing. Notably, IQM Garnet is also currently a part of IQM's cloud offering, IQM Resonance [18].

The performance level described above is representative to what is achieved with today's integrated superconducting quantum computing systems. In the short term, we anticipate improvements through control optimization, improvement of coherence, and incremental improvements in the system and QPU. Soon we will integrate recently reported control optimization [16], and aim to improve relaxation time $T_1$ from the order of $40\,\mu s$ to above $100\,\mu s$ which we have already demonstrated in test devices [13].